\documentclass[12pt,preprint]{aastex}

\newcommand{\bb}{\begin{equation}}
\newcommand{\ee}{\end{equation}}

\shorttitle{Sunspot Oscillations}
\shortauthors{Rajaguru et al.}

\begin{document}

\title{Radiative transfer effects on Doppler measurements as sources of surface effects in 
sunspot seismology}
\author{S.P. Rajaguru\altaffilmark{1}, K. Sankarasubramanian\altaffilmark{2}, 
R. Wachter\altaffilmark{1} and P.H. Scherrer\altaffilmark{1}}
\altaffiltext{1}{W.W. Hansen Experimental Physics Laboratory, Stanford University, Stanford CA 94305}
\altaffiltext{2}{Space Astronomy and Instrumentation Division, ISRO Satellite Centre, Bangalore, India} 

\begin{abstract} 
We show that the use of Doppler shifts of Zeeman sensitive spectral lines to observe waves
in sunspots is subject to measurement specific phase shifts arising from, 
(i) altered height range of spectral line formation and the propagating 
character of p mode waves in penumbrae, and (ii) Zeeman broadening and splitting.
We also show that these phase shifts depend on wave frequencies, strengths and line of sight 
inclination of magnetic field, and the polarization state used for Doppler measurements.
We discuss how these phase shifts could contribute to local helioseismic measurements of
'surface effects' in sunspot seismology.
\end{abstract}

\section{Introduction}
\label{sec:intro}

The solar oscillations are measured with better signal-to-noise from Doppler shifts of a spectral line than
intensity fluctuations. Existence of reliable algorithms to estimate Doppler shifts
from line intensities (filtergrams) scanned over a few locations across a line (MDI; \citet{scherreretal95}),
or a fast Fourier tachometric scanning of a line (GONG; \citet{harveyetal94}), makes possible deriving velocity
images of solar surface frequent enough to study the solar oscillations. 
Above methods of estimating line of sight velocities from Doppler shifts assume
rigid displacements of a spectral line in wavelength in response to material motions.
However, gradients along line of sight in solar velocity and other physical quantities
introduce asymmetries in line profile leading to Doppler cross-talk of signals from
different layers in the atmosphere. In quiet Sun, the five minute band p mode oscillation 
signals are evanescent at observation heights and hence oscillation phases are only minimally
affected by height gradients.
In active regions, not only the vastly different thermal 
and magnetic structure complicate Doppler shift measurements through changes in line profile shapes but also 
altered character of waves in line forming layers.

In this Letter, using a temporal sequence of spectropolarimetric observations of a sunspot
we demonstrate: (i) propagating character of p modes in the penumbra and 
extended height range of spectral line formation cause phases of waves to depend strongly
on intensity level within a line from where Doppler velocities are measured,
(ii) sampling Zeeman split portions near the line core causes 
filtergraph (MDI/SOHO) or bisector methods
for Doppler velocity estimation to introduce spurious phase shifts in the waves measured.
Wave propagation signatures, in the first case above, which have recently been studied
in chromospheric lines \citep{finsterleetal04}, are also shown to differ
in the Doppler shifts of Stokes I and polarized components (circular and linear).
We discuss how the above findings are related to recent
local helioseismic measurements of 'surface effects' \citep{schunkeretal05,zhaoetal06}. 

\section{Data and Analysis Methods}
The observations were carried out using the Advanced Stokes Polarimeter (ASP) at the
Dunn Solar Telescope of the National Solar Observatory at Sacramento Peak, Sunspot, New Mexico. 
A medium sized sunspot (NOAA AR0750, diameter $\approx$ 16 Mm) located close to disk center
was observed on April 12, 2005, in mid-photospheric spectral line Ni {\sc i} ($\lambda$=6768 \AA, 
used by MDI/SOHO and GONG).
We have made a temporal sequence of Stokes vectors of this line by scanning
the sunspot over consecutive (in $x$-direction) 14 slit positions oriented N-S ($y$-direction)
in one minute and repeating it for about 100 minutes (with a solar image tracker fixing the field of view).
The N-S extent of slit covers about 64 Mm with a resolution of 0.375$^{"}$ with the spot at the center
and, with a slit width of 0.6$^{"}$, the 14 slit positions (x-direction) cover mainly umbral region.

We do Milne-Eddington (M-E) inversions of Stokes vectors ($I,Q,U,V$'s) to obtain line of sight velocities
$v(x,y,t)$, total magnetic field strengths $B(x,y,t)$, line of sight inclination $\gamma(x,y,t)$ and 
azimuth $\psi(x,y,t)$ of magnetic field among other standard inverted quantities \citep{sku-lites87}. 
The M-E inversion procedure is weighted towards polarized component within magnetic 
field and provides velocity estimates from the line core position which is used as a free parameter
\citep{sku-lites87,jefferiesetal89}.
Since we want to examine how wave phases change with height within line
formation layers, how they are seen in polarized and unpolarized light,  and how they contribute to 
local helioseismology measurements of 'surface magnetic effects', we also obtain line of sight velocities
from the following methods: (i) Doppler shifts of bisector points at different intensity levels within $I$
\citep{keil-yackovich81,cavallinietal85,rimmele95},
and those within circular (average of $I+V$ and $I-V$ velocities, hereafter denoted CP) and linear 
($I+Q$, denoted LP1) polarized components \citep{sankar-rimmele02,dtiniesta03},
and (ii) the two Doppler filtergram measurements in circular (CP) and linear (LP1) polarized 
components employed by MDI/SOHO \citep{scherreretal95}.

Phase shifts between any two velocity signals, $v_{1}(t)$ and $v_{2}(t)$, are calculated according to,
\begin{equation}
\phi_{1,2}(\omega)=tan^{-1}\left (\frac{Im[v_{1}(\omega)v^*_{2}(\omega)]}{Re[v_{1}(\omega)v^*_{2}(\omega)]}
\right) ,
\label{eq:phase}
\end{equation}
where $v_{1}(\omega)$ and $v_{2}(\omega)$ are the Fourier transforms of the two time series. In the
above convention, positive values for $\phi_{1,2}$ at positive $\omega$ mean $v_{1}(t)$ is 
advanced in phase with respect to $v_{2}(t)$. 
Our aim here is to study how $\phi$ between velocities from different bisector levels as well as that between 
different measurements vary over the sunspot as a function of position and $\omega$, i.e. $\phi(x,y,z,\omega)$,
due to propagating waves and altered line profile shapes (Zeeman splitting). The vertical
co-ordinate $z$ represents height within line forming layers and we use intensity levels 
(i.e., bisector points) within the line to scan it. Based on the plane-parallel VAL-C model of quiet solar
atmosphere, the Ni {\sc i} line formation is reported to span a height range of 
18 km (continuum) -- 288 km (line core) ($z$=0 being referenced to continuum optical 
depth $\tau_{c}$=1) \citep{nortonetal06,bruls93}. We do not attempt full Stokes profiles inversions 
to determine $z$ dependences of $\phi$; 
this is not only computationaly expensive (as we have too many profiles from the $x,y,t$
scans of Stokes vector) but also lead to noisier signals than those from direct
bisector analyses. In quiet-Sun region 
the main p mode band acoustic waves ($\omega < \omega_{ac} \approx $5.3 mHz, the acoustic 
cut-off frequency) are largely reflected down into the solar interior, and hence 
$\phi(x,y,z,\omega) \approx$ constant = 0; at $\omega > \omega_{ac}$, however, because  
of wave propagation, $\phi(x,y,z,\omega) \approx \phi(z,\omega) \neq$ 0. 
Within the sunspot, we study $x$ and $y$ dependences of $\phi$ using 
temporal averages of M-E inverted $B(x,y,t)$ and $\gamma(x,y,t)$ shown in Figure 1.

\section{Surface magnetism effects due to propagating waves and Zeeman splitting}

In order to analyse phase shifts due to the propagating nature of waves,
we calculate bisector velocities from 9 equally spaced intensity levels within $I$ 
and CP profiles, $I_{1}$=0.1$I_{c}$, $I_{2}$=0.2$I_{c}$,..., $I_{9}$=0.9$I_{c}$, where $I_{c}$ is
the continuum level. Using the time series of velocities from different 
bisector levels, we calculate phase shifts between them using Equation \ref{eq:phase}.
When the line is Zeeman split, formation heights
of bisector levels of $I$ profiles near the core are significantly altered and any asymmetries in 
depths of Zeeman components further corrupt the velocity signals. Bisector points of individual 
CP components are largely free of the above uncertainties. In view of the
above, we use M-E inverted velocities as representative of the highest layer within line formation
region, and we show that this is indeed the case by comparing them with CP bisector velocities 
from near the core for the Ni {\sc i} line.

The phase shifts $\phi(x,y,z,\omega)$
exhibit random fluctuations over space as well as in $\omega$ due to the inherent stochastic 
nature of the solar velocity field and due to noise from various sources. Fluctuations over $\omega$
are reduced by taking median values of $\phi$ over the p mode band (2 - 5 mHz) or
over bands of 1 mHz width centered at 3, 4 and 5 mHz (to study any frequency dependence). Signals 
over space are studied using the temporal averages $B(x,y)$ and $\gamma(x,y)$ (Figure 1):
we average the phase shifts over all pixels that fall within a 25 G window
of $B$ or 2$^{\circ}$ of $\gamma$ and assign the average to all those pixels (spatially) and to the
central values of $B$ and $\gamma$. The resulting spatial maps of $\phi$ are essentially one dimensional
functions of $B$ or $\gamma$ with any azimuthal variations due to such variations in $B$ or $\gamma$ 
averaged out. In Figure 2, we plot $\phi$ between mid-level 
($I_{5}=0.5I_{c}$) bisector velocities and those from other levels: panels $a$ and $b$ 
show median $\phi^{CP}_{5,i}$(i=1,3,7,9) (superscripts identify the profiles used and subscripts 
the pair of levels that the phase shifts come from) over the full p mode band as a function of 
$B$ and $\gamma$, respectively, 
panel $c$ shows $\phi^{CP}_{5,1}$ and $\phi^{CP}_{5,9}$ as a function of $B$ for different frequency bands,
and panel $d$ compares $\phi^{CP}$ and $\phi^{I}$ between levels (5,1) and (5,9) as a function of $B$ for
the full p mode band. For clarity, we have plotted error bars only at representative values of
$B$ or $\gamma$. The error bars correspond to standard deviations of $\phi$ over the pixels that fall within
the bin sizes of $B$ or $\gamma$. The spatial maps that are obtained after binning over 25 G windows in line
of sight magnetic field $B_{z}=B cos(\gamma)$ are shown in Figure 3:
in panels $a$ and $b$, respectively, are $\phi^{I}_{5,i}$ and $\phi^{CP}_{5,i}$, where $i=$M-E,1,2 and 9, 
in panel $c$ are $\phi^{I,CP}$ that are between same level bisector velocities from CP and $I$ profiles, 
and in panel $d$ are $\phi^{CP}_{1,9}$ and $\phi^{I}_{1,9}$ in the 2.5 - 3.5 mHz and 3.5 - 4.5 mHz frequency bands.
Changes in $\phi$ against $B$ or $B_{z}$ and the bisector intensity levels, in Figures 2 and 3,
clearly show that these are due to propagating waves, even within the main p mode band, and that propagation is
mainly confined to penumbral region.
In quiet Sun pixels ($B <$ 200 G), $\phi$ due to wave propagation into higher layers at $\omega > \omega_{ac}$, 
though much smaller than that seen in the penumbra, is noticeable in panels $c$ and $d$ of Figure 2.
Both Stokes $I$ and CP profiles' velocities show propagating waves but the signals are larger in CP
bisector velocities indicating that propagation is mostly in magnetized channels (panel $d$ of Figure 2 and
panels $a$ and $b$ of Figure 3). Phase shifts between
same level bisector velocities from $I$ and CP profiles, shown in panel $c$ of Figure 3, further substantiates
the above inference.
The $\phi^{I,CP}$ from wing level (7 and 9, the right two panels in panel $c$ of Figure 3) 
bisector velocities are very small, suggesting that differences in formation heights of wings of $I$ and
CP profiles is small. 

The above results show that if we have to
minimize contributions from propagating waves we should restrict Doppler measurements to wings. 
In panel $d$ of Figure 2 and panel $a$ of Figure 3, it is seen that the Stokes $I$ velocities introduce spurious 
advanced phases within the umbral and umbral - penumbral boundary region, when bisector velocities from
near the core are involved: the fallen or collapsed $\phi^{I}_{5,1}$ plotted as red triangles in panel $d$ of Figure 2,
and the scrambled phase signals seen in the middle two maps of panel $a$ in Figure 3. We attribute the above
to changes in $I$ profile shapes due to Zeeman splitting, as these signals are not present or weak in CP
measurements. We confirm this using the M-E inverted velocities, which model the Zeeman splitting and retrieve
velocities; the left most map in panel $a$ of Figure 3 show that $\phi^{I}_{5,ME}$ is free of the above discussed
artifacts, and also matches well with map of $\phi^{CP}_{5,ME}$ in panel $b$ of Figure 3. 
The similar values and structure of $\phi^{CP}_{5,1}$ and $\phi^{CP}_{5,2}$ to that of $\phi^{CP}_{5,ME}$ further
clarifies that wave propagation is confined to magnetized medium. Interestingly, the M-E velocities
show negative changes for $\phi^{I}_{5,ME}$ and $\phi^{CP}_{5,ME}$ in the outer penumbra (panels
$a$, $b$ of Figure 3), which we identify as a possible signature of downward propagating waves along
field lines that return back to the photosphere. 

The MDI measurements \citep{scherreretal95} use CP and LP1 profiles: 
most full-disk observations are based on LP1
and almost all helioseismic holography results \citep{braun-lindsey99,braunetal04,schunkeretal05,
lindseyandbraun05a,lindseyandbraun05b} on
surface magnetism effects are based on these data; on the other hand, most time-distance helioseismology
studies of sub-surface structure and flows \citep{sashaetal00,zhaoetal01} and the surface
magnetism effects \citep{zhaoetal06} have used the MDI CP hi-res data. Here, we have derived
Doppler velocities from both of these MDI measurements by executing the MDI algorithms (which the
onboard processor of MDI uses). In quiet Sun, i.e. when there are no strong magnetic fields to 
modify the line profile through Zeeman effect, both of these MDI measurements reduce to deriving
Doppler velocities from unpolarized Stokes $I$ and hence are the same. However, as demonstrated in Figures 2 and 3,
propagating waves introduce phase shifts, and to estimate how much of these are present in MDI measurements
we have calculated phase shifts between MDI CP and LP1 velocities and wing level (level 7) bisector
velocities from $I$, denoted respectively as $\phi^{CP}_{MDI}$ and $\phi^{LP1}_{MDI}$. 
The results are in Figure 4: panel $a$ shows $\phi^{LP1}_{MDI}$ 
and panel $b$ shows $\phi^{CP}_{MDI}$ over bands of 1 mHz width centered at frequencies of 
3, 4 and 5 mHz and also the full p mode band (2 - 5 mHz). We found from bisector velocities from near the
core that Zeeman splitting causes an advancement of phases within umbra (panel $a$ and $d$ in Figure 3);
since LP1 profile within umbra is the same as $I$ (because $\gamma$ is close to zero), similar advanced
phases are expected from MDI LP1 measurements. This is clear in panel $a$ of Figure 4.
In summary, for the medium size spot observed, surface effects due to wave propagation in the penumbra 
cause a phase lag in the range of 5 - 10 seconds in both MDI LP1 and CP measurements and the Zeeman
split profiles cause positive phase shifts of the order of 8 seconds in umbra in MDI LP1 measurements.

\section{Discussions and conclusion}
Recent studies in sunspot seismology \citep{schunkeretal05,lindseyandbraun05a,lindseyandbraun05b,zhaoetal06} 
have recognised the role of 'surface magnetic effects' that contribute to seismically measured
phases or travel times of acoustic waves. 
We have shown that phases of acoustic waves observed within sunspots using Doppler shifts
of a spectral line have signatures of physical changes that waves undergo within line forming
layers as well as of systematics in Doppler measurements induced by Zeeman split profiles.
We have also shown that wave propagation effects and systematics from 
split portions of the line profiles are minimal when Doppler velocity measurements are restricted 
to wings of a spectral line.
Since a seismologically correct determination of phase shifts
require separating the above surface effects in the measurements, our latter results 
above show that such a correction can be carried out in the observation procedure itself: within sunspots,
the Doppler shift measurements could be restricted to wings of spectral lines. In the case of MDI/SOHO,
the onboard algorithm combines all the filtergrams
to determine Doppler velocities. Since the filter positions
are fixed with respect to the central wavelength, when
the line is wider due to Zeeman splitting, there is a systematic sampling of split 
core region of the profiles and hence higher atmospheric layers. 
This causes phase lags in waves from penumbral region and spurious advanced
phases within umbra when LP1 profile is used (most full-disk measurements). 
Correcting the MDI measurements for such surface effects, 
hence, requires maps such as the ones derived here (Figure 4). 
A detailed 'calibration' of phase shifts due to surface effects against
$B$ or $\gamma$ require similar studies involving spots of various sizes.
 
We have not assessed how the surface effects studied here would depend on positions across the solar disk.
However, the present finding that propagating
waves in penumbra cause a phase shift depending on height of observation
points to similar effects when there are line of sight changes in optical depth. Such changes
are expected for sunspots located away from disk center, due to the Wilson depression, and
this combined with the field line alligned wave propagation can introduce opposite changes,
in measured phases, in the limb side (deeper level of observation) and center side (higher level
of observation) penumbrae. We suggest that such effects on the
phase shifts are a likely contributor to the azimuthally varying 'inclined magnetic field 
effect' of \citet{schunkeretal05} and \citet{zhaoetal06}. The magnitude of surface effect signal
that we have, for the medium sized spot, falls in the range of  5 - 15 seconds (Figures 2, 3 and 4)
and it points to larger phase shifts for larger spots. We note 
that, for fairly large sunspots, the magnitude of travel time perturbations measured in 
time-distance helioseismic measurements \citep{duvalletal96,sashaetal00,zhaoetal01}, 
as well as the vantage dependent control correlation phases \citep{schunkeretal05}, 
fall in the range of 30 - 60 seconds. Our results
on surface effect signal due to Zeeman splitting in umbral region (Figures 3 and 4) call
for a careful study of active regions in one or more magnetically insensitive photospheric lines.

\acknowledgments
This work is supported by NASA grants NNG05GH14G to $SOHO$ MDI project and NNG05GM85G to Living With a Star 
(LWS) program at Stanford University.

\clearpage

\begin{figure}
\epsscale{0.8}
\plotone{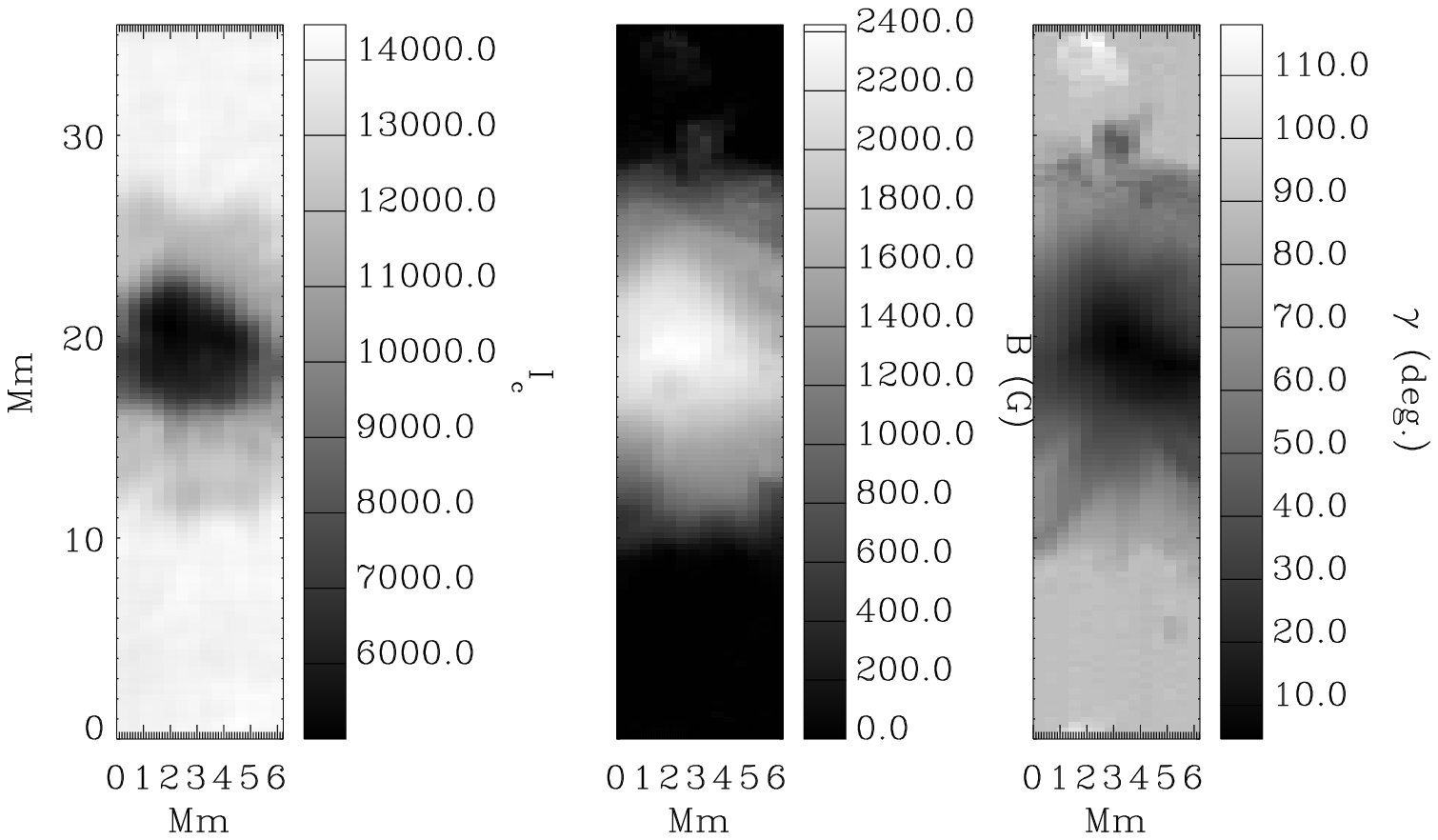}
\caption{Maps of continuum intensity $I_{c}$ and temporal averages of M-E inverted $B$ and $\gamma$ 
over the sunspot from Ni {\sc i} observations.}
\label{fig:1}
\end{figure}

\clearpage

\begin{figure}
\epsscale{1.0}
\plottwo{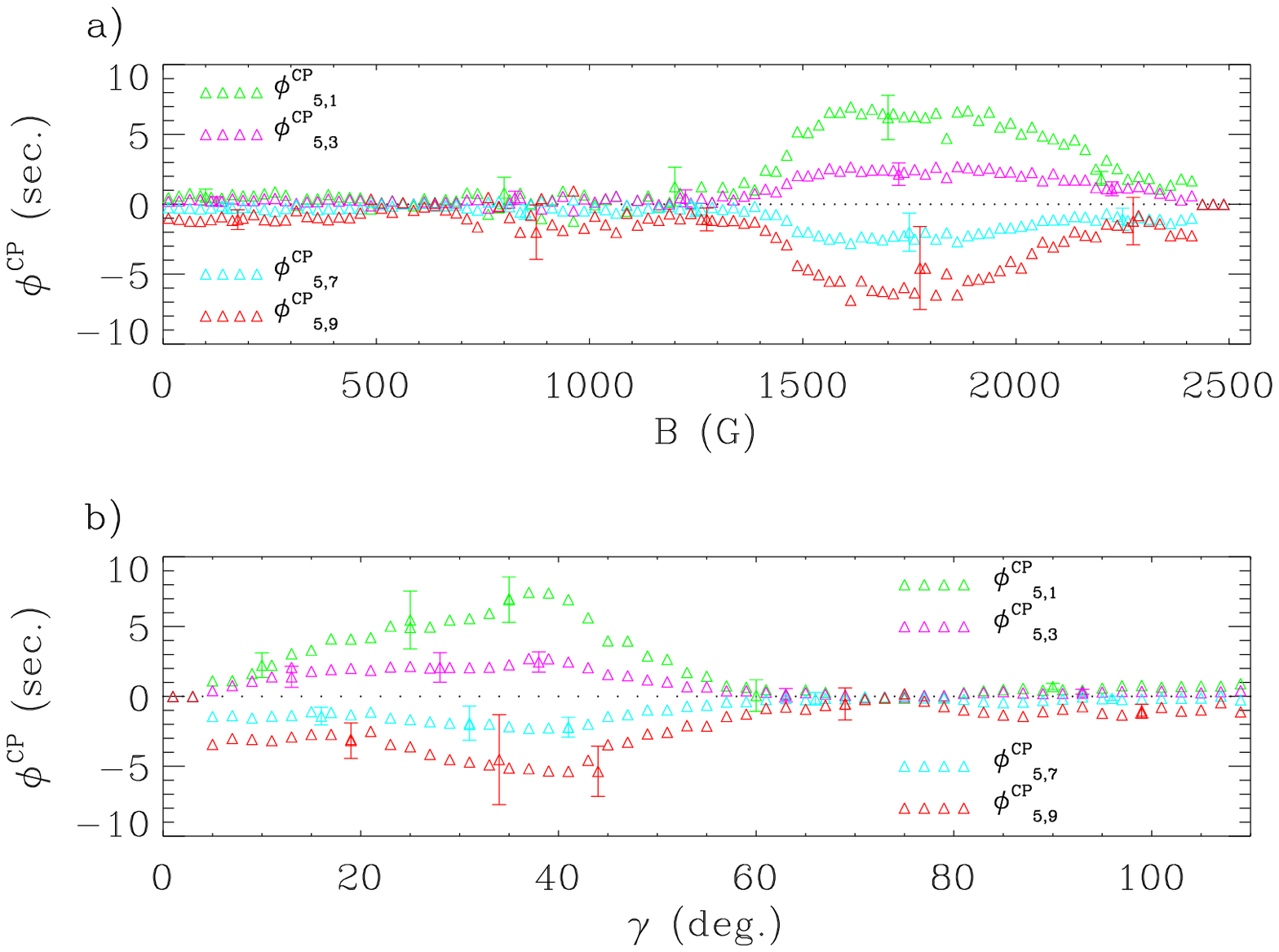}{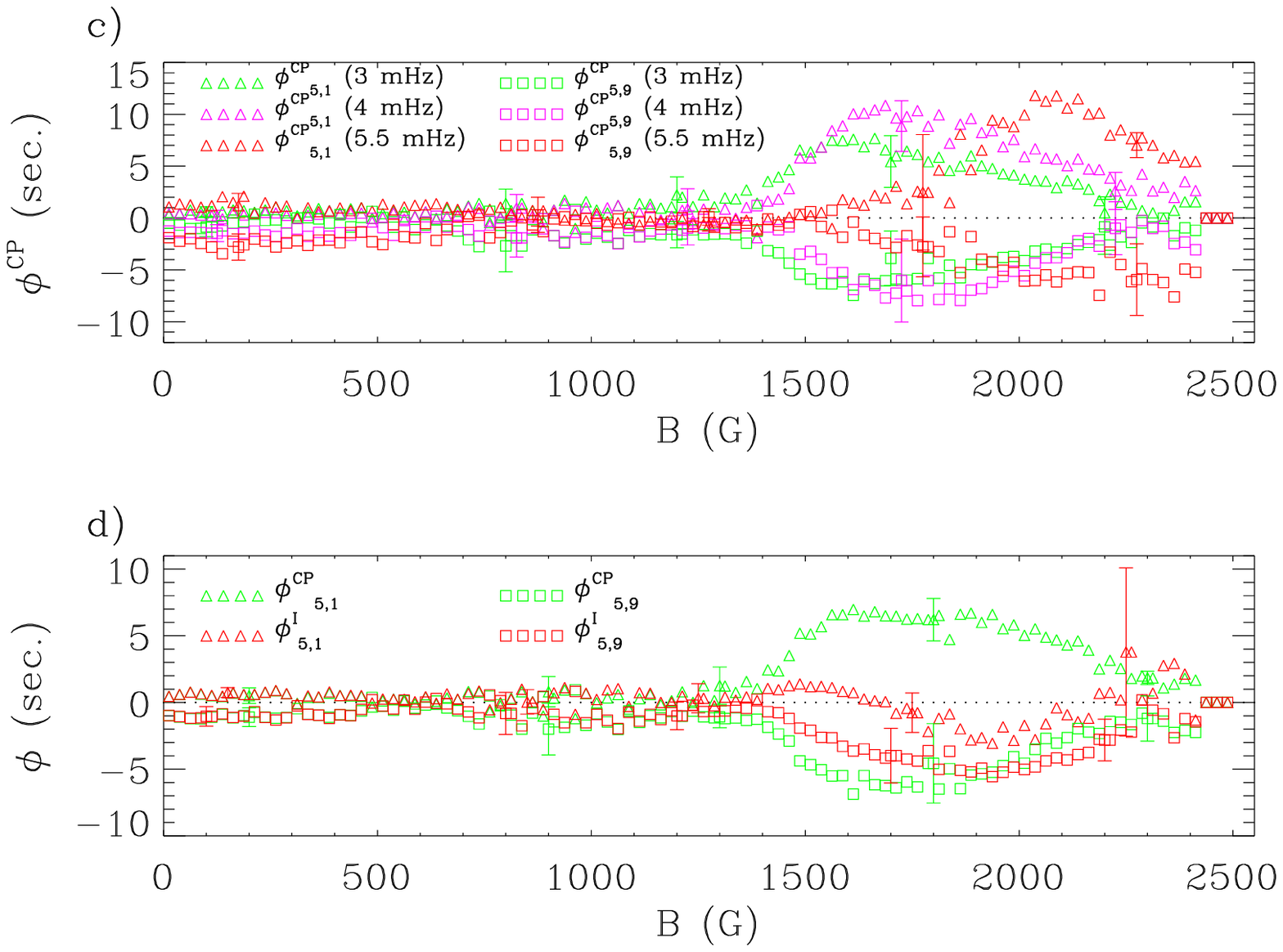}
\caption{Phase shifts between velocities from different bisector levels of $CP$ and $I$ profiles of
Ni {\sc i} line: panels $a$ and $b$ show $\phi^{CP}$ against $B$ and $\gamma$, respectively,
for the full p mode band (2-5 mHz), panel $c$ shows $\phi^{CP}_{5,1}$ and $\phi^{CP}_{5,9}$ over different
frequency bands, and panel $d$ compares $\phi^{CP}$ and $\phi^{I}$ from the bisector levels (5,1) and (5,9)
for the full p mode band; see text for further details}
\label{fig:2}
\end{figure}

\clearpage

\begin{figure}
\figurenum{3}
\plottwo{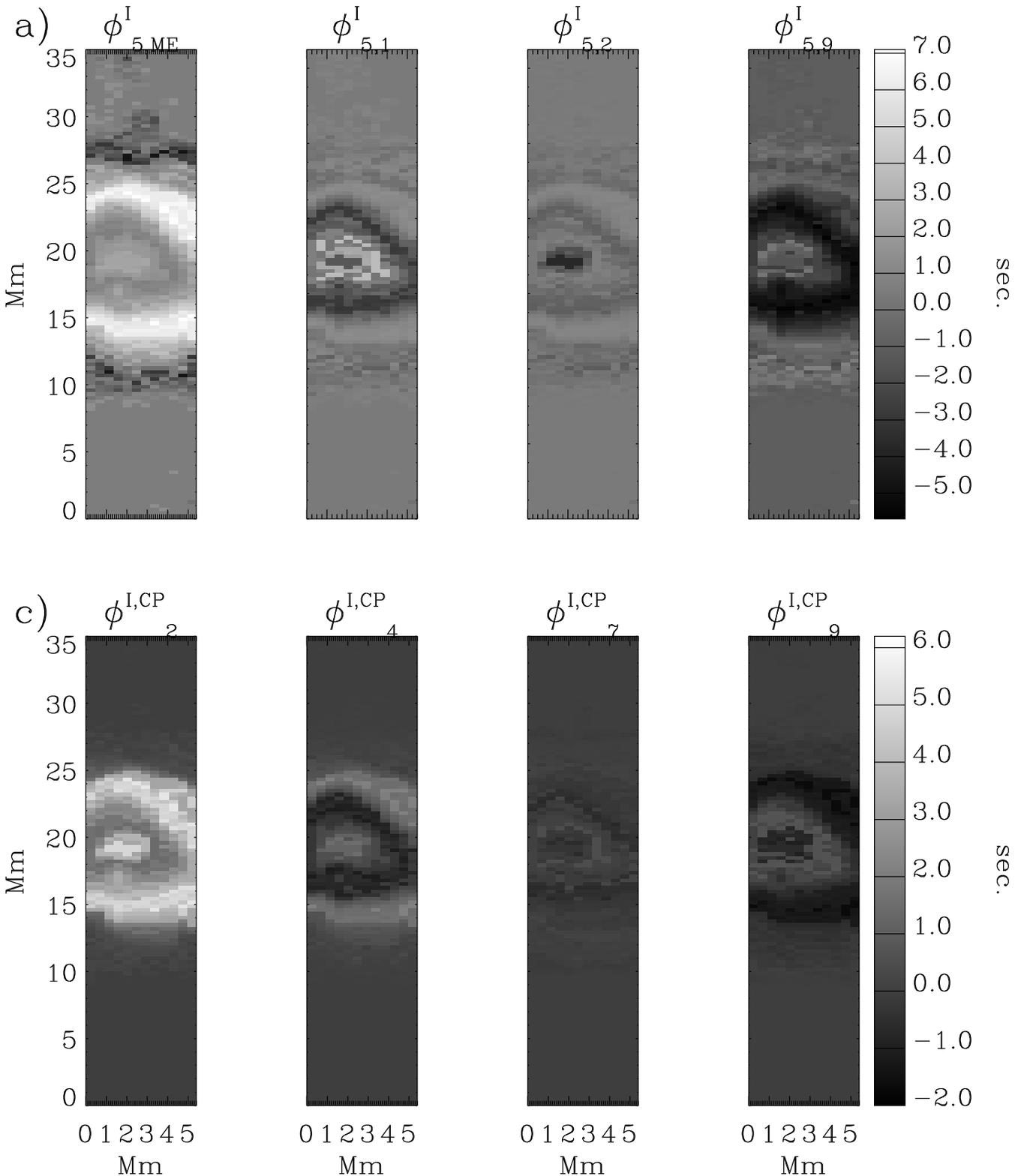}{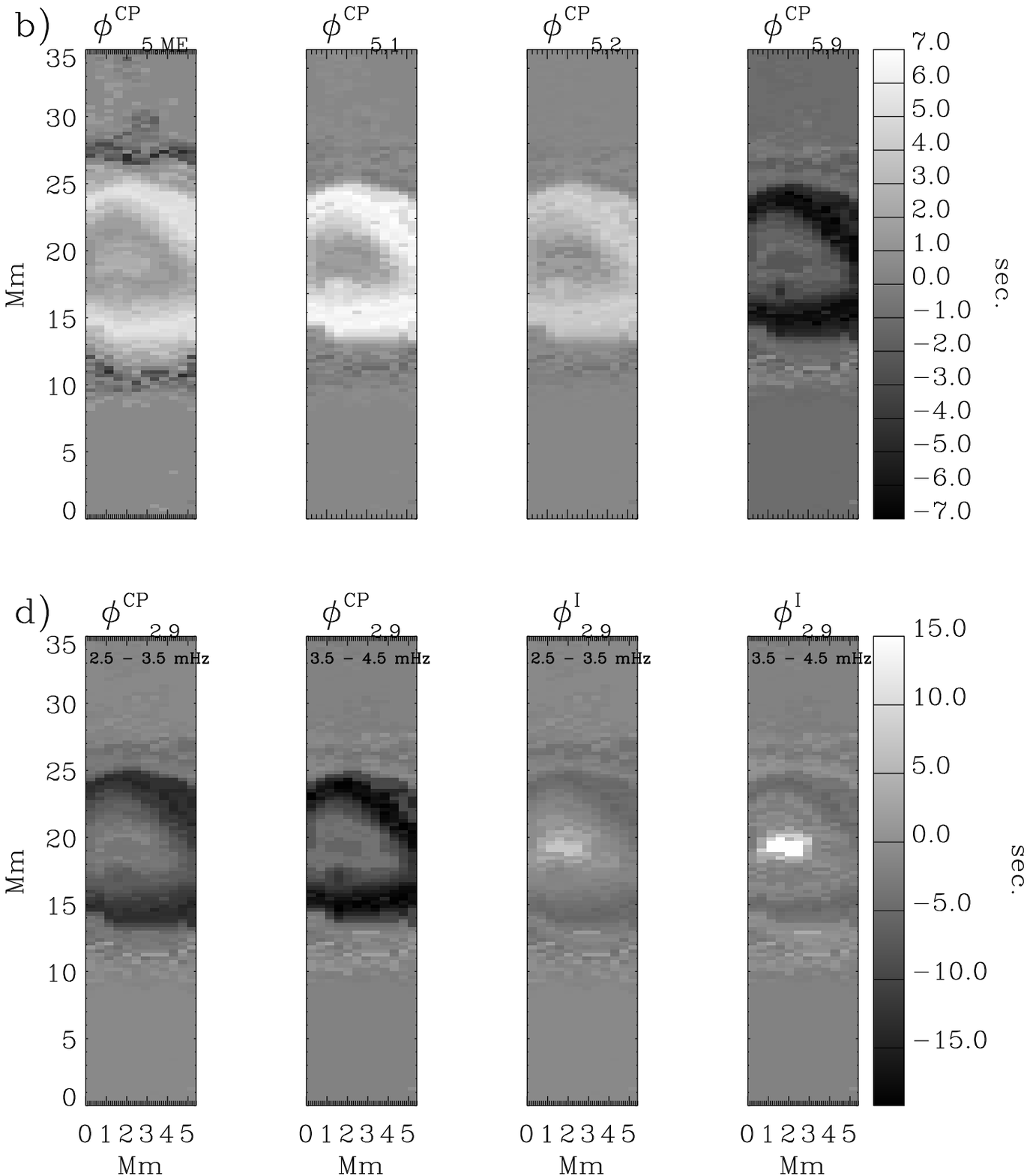}
\caption{Maps of phase shifts of waves observed using Ni {\sc i} line; see text for details}
\label{fig:3}
\end{figure}

\clearpage

\begin{figure}
\figurenum{4}
\epsscale{0.9}
\plotone{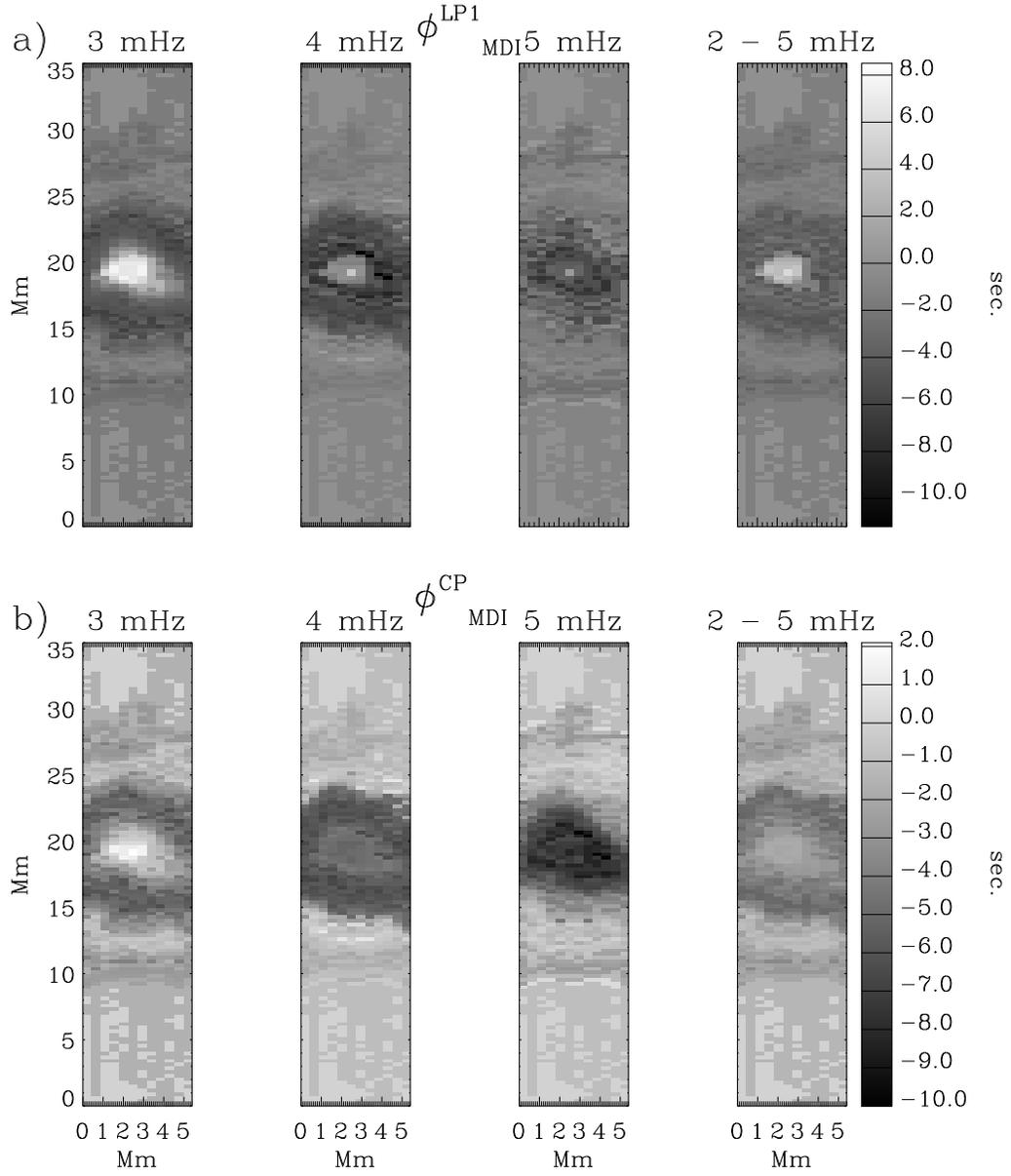}
\caption{Maps of phase shifts due to 'surface magnetic effects' in MDI linear (LP1, full disk) and circular polarization
(CP, hi-res) measurements.}
\label{fig:4}
\end{figure}

\end{document}